\documentclass[3p]{elsarticle}

\makeatletter
\def\ps@pprintTitle{%
 \let\@oddhead\@empty
 \let\@evenhead\@empty
 \def\@oddfoot{}%
 \let\@evenfoot\@oddfoot}
\makeatother

\usepackage{lineno,hyperref}
\modulolinenumbers[5]

\usepackage{xcolor}
\usepackage{newtxmath}
\usepackage{hyperref}
\usepackage{cleveref}
\usepackage{upgreek}
\usepackage{subcaption}    
\usepackage{setspace}

\DeclareMathOperator*{\argminA}{arg\,min}

\begin{document}
\newcommand{\bb}{\boldsymbol}
\newcommand{\mb}{\mathbf}
\DeclareRobustCommand*{\drv}{\mathop{}\!\mathrm{d}}
\begin{frontmatter}

\title{Symmetry-Aware Autoencoders: \\ s-PCA and s-NLPCA}

\author[imperial]{Simon Kneer\corref{mycorrespondingauthor}}
\cortext[mycorrespondingauthor]{Corresponding author}
\ead{kneer@pks.mpg.de}
\author[cnrs]{Taraneh Sayadi}
\author[onera]{Denis Sipp}
\author[imperial_2]{Peter Schmid}
\author[imperial]{Georgios Rigas}

\address[imperial]{Department of Aeronautics, Imperial College London, London SW7 2AZ, UK}
\address[cnrs]{Jean le Rond d’Alembert Institute, CNRS/Sorbonne University, Paris, France}
\address[onera]{DAAA, ONERA, Université Paris Saclay, 8 rue des Vertugadins, 92190 Meudon, France}
\address[imperial_2]{Department of Mathematics, Imperial College London, London SW7 2AZ, UK}

\begin{abstract}
Nonlinear principal component analysis (NLPCA) via autoencoders has attracted attention in the dynamical systems community due to its larger compression rate when compared to linear principal component analysis (PCA). These model reduction methods experience an increase in the dimensionality of the latent space when applied to datasets that exhibit invariant samples due to the presence of symmetries. In this study, we introduce a novel machine learning embedding for autoencoders, which uses Siamese networks and spatial transformer networks to account for discrete and continuous symmetries, respectively. The Siamese branches autonomously find a fundamental domain to which all samples are transformed, without introducing human bias. The spatial transformer network discovers the optimal slicing template for continuous translations so that invariant samples are aligned in the homogeneous direction. Thus, the proposed symmetry-aware autoencoder is invariant to predetermined input transformations.  This embedding can be employed with both linear and nonlinear reduction methods, which we term symmetry-aware PCA (s-PCA) and symmetry-aware NLPCA (s-NLPCA). We apply the proposed framework to the Kolmogorov flow to showcase the capabilities for a system exhibiting both a continuous symmetry as well as discrete symmetries.
\end{abstract}


\end{frontmatter}

\section{Introduction}
\label{sec:intro}
Modal decomposition methods have been widely used to further the understanding of complex physical processes for decades. In the field of fluid dynamics, ever since principal component analysis (PCA), also known as proper orthogonal decomposition (POD) was introduced by \citet{lumley1967}, it has been a staple tool for decomposing flows into relevant structures, since it gives an optimal linear representation of the flow in terms of the $L_2$-norm.
\par
The connection between {\color{black} neural networks} and PCA was first drawn by \citet{BALDI1989}. In their study, an autoencoder (AE) that is constructed as a linear multilayer perceptron (MLP) was employed to show its equivalence to PCA. An AE is a compilation of two networks, an encoder and a decoder, in series. Thus, an input can be reduced or encoded to a latent space, which is analogous to the space of amplitude coefficients for PCA, and afterwards reconstructed or decoded into the original input space. \citet{plaut2018} was able to show that the space that is spanned by a number of PCA components and a linear AE with the same amount of components are identical. The components for the latter method are not orthogonal by default, unlike for PCA, but can be approximately orthogonalized by employing a singular value decomposition on the weights after training \citep{plaut2018}. However, PCA is only equivalent to linear AEs without biases, which are only a small subset of such networks.
In a more general setting, an AE is deployed in combination with activation functions that allow the network to learn nonlinear relations between the input and the output. This nonlinear PCA (NLPCA) for fluid dynamics was first employed by \citet{Milano2002} and showed great promise in reducing the $L_2$-reconstruction error for the randomly forced Burgers equation and turbulent channel flow when comparing to PCA. However, as of now, there is no unified way to visualize NLPCA modes, since the temporal and spatial behaviour are nonlinearly coupled. Different promising approaches have been suggested. \Citet{fukami2020} force the network to be additive in the final output space, which allows for a comprehensive way to investigate the spatial structures of each output channel. \citet{Page_2021} performed a latent Fourier analysis, that is able to extract patterns based on a filtering of the latent wavenumbers. 
\par
Physical equations, such as the Navier-Stokes equations, are equivariant under symmetry transformations, leading to invariant solutions that are identical up to a prescribed transformation function. The presence of invariant solutions poses an issue for model reduction techniques, since conventional methods cannot find the connection between them and, thus, encode them into different regions of the latent space. This leads to perceived higher dimension of the latent space since the optimization towards an optimal basis becomes harder. 
Additionally, for linear methods, where the focus usually lies in the extraction of  the modal shapes to gain an insight into the flow structures, these invariant solutions can morph the modes away from physical structures. {\color{black}In fact, given a large enough dataset of an ergodic system,
a necessary condition is that the space spanned by PCA modes is invariant with respect to the same symmetry operations the dynamical system is equivariant under. Observed instantaneous events are not bound by this condition, however, creating a split between observed states and the modal basis, see \citet{holmes2012}}.
\par
In the case of PCA, this issue has long been observed for periodic boundaries, which create a group of translationally invariant solutions, $\bb{T}(s):u(x)\rightarrow u(x+s)$. For datasets of ergodic turbulent fluid flows on domains with at least one homogeneous direction, that are both statistically stationary and have reached statistical convergence with respect to the translational symmetries, the modal representation will become the Fourier basis \citep{SIROVICH1987}. In the case of NLPCA these symmetry copies show themselves in the larger dispersion of the variables in the reduced space, see e.g. \citet{Page_2021}. Different methods to circumvent these issues have been proposed in the past. \Citet{mendible2020} fit prescribed template functions to their data in order to extract an exact definition of the temporal evolution of the wave speed of travelling wave problems. Another method called template matching, see \citet{Rowley_2003} and \citet{holmes2012}, aims to remove continuous symmetries by matching a template, chosen {\it a priori}, with the flow field and translating each snapshot by an offset from the template. Another method, known as the method of slices \citep{budanur2015}, aims to remove translationally invariant solutions by {\color{black}{mapping each of the states to a slice template. Such a slice template can for example be the first fourier mode, which leads to a reduction of the continuous symmetry by rotating each state so that its first Fourier mode is purely real}}. In combination with NLPCA, this method has successfully been applied by e.g. \citet{Linot2020}, who also included temporal advancement in the reduced space. Using the first Fourier mode as a slice template is only valid when the amplitude of this mode is sufficiently large, since quasi discontinuous jumps can appear otherwise. This problem is known to arise in turbulent channel flow \citep{marensi2021} and can be circumvented by introducing a handcrafted modulation of the Fourier basis in the inhomogeneous direction.
\par
Discrete symmetry groups, $\bb{G}$, such as the reflectional symmetry in a cylinder wake, $\bb{R}:(u,v)(x,y)\rightarrow (u,-v)(x,-y)$, are of importance as well. Invariant polynomial bases can be constructed to unify different discrete symmetries while keeping the dynamics smooth, but these have to be tailored to each problem, and in many cases no such bases have been found as of now \citep{budanur2015exact}, \cite{Budanur2016}. If one disregards the smoothness of dynamics in preference of remaining in the space of physical observables, one can try to find a fundamental domain to map the states to, however, such fundamental domains can be hard to find and carry with them an inherent human bias, since there is infinitely many possible fundamental domains.
\par
In this work we propose a machine learning method, called symmetry-aware principal component analysis (s-PCA), in order to efficiently reduce the complexity of dynamical systems, exhibiting discrete or continuous invariant solutions.
Discrete symmetries are removed by siamese branches, i.e. multiple autoencoders in parallel. This method is closely related to the one presented by \cite{mehr2018}, however, differs in the fact that we do not combine our branches in the latent space. Hence, our method discovers a fundamental domain to which all states are mapped without user bias. To remove continuous symmetries we employ spatial transformer networks \cite{jaderberg2015}, which discover a single shift parameter for the translated direction. Since translations are linear transformations, we apply linear spatial transformer networks. These then find an optimal slicing template without human intervention.
\par
We apply this symmetry reduction to both the linear (s-PCA) and nonlinear (s-NLPCA) reduction methods, creating a higher degree of compression as well as improving the interpretability of the resulting modes.
\par
After an in-depth analysis of our proposed method, investigations into the effectiveness of the above methodology are performed on the two-dimensional Kolmogorov flow with $n=4$. Kolmogorov flow is equivariant under the group of continuous translations, $\bb{T}(s)$, a discrete shift-and-reflect symmetry, $\bb{S}$: $\omega(x,y)\rightarrow \omega(-x,y+\frac{\pi}{n})$, and a discrete rotation through $\pi$, $\bb{P}$: $\omega(x,y)\rightarrow \omega(-x,-y)$. 
\section{Background}
In this paper we employ multiple model reduction methods to address difficulties arising from invariant solutions within a dataset, and propose remedies to circumvent these problems. On one hand, a dataset can be reduced with a linear parametric model, which can be constructed with the analytical method of PCA, and its neural network equivalent using linear autoencoders. As has been shown before \citep{BALDI1989}, both methods span the same subspace independent of the construction method, however, using the latter orthogonality of the modes is not given without further modification. On the other hand, there is the nonlinear generalization of PCA, NLPCA, which can be constructed using nonlinear autoencoders. Both linear and nonlinear methods can be embedded into the symmetry-reducing framework we have developed, leading to linear (s-PCA) and nonlinear symmetry-aware PCA (s-NLPCA), respectively. First, an overview of different invariant solutions commonly found in physical problems is provided. 
\subsection{Invariant solutions}
We call a solution, $\bb{X}(\bb{\xi},t)$, to a governing system of equations dependent on the spatial coordinates $\bb{\xi}$ and time $t$, invariant under a given symmetry operation, $\bb{S}$, if $\bb{SX}(\bb{\xi},t)$ is also a solution to the governing equations. We can split $\bb{S}$ into two groups: continuous and discrete symmetry groups. 
\par
Continuous symmetry groups, $\bb{F}(s)$, are based on continuous operators that can be applied to a solution. These operators depend on continuous transformation parameters, denoted by $s$. In the case of an incompressible two-dimensional channel flow, $(u,v,p)(x,y)$, with periodic boundary conditions in $x$, we observe such a symmetry group: the group of continuous translations, $\bb{T}(s)$. Any solution $(u,v,p)(x+s,y)$ is a solution to the Navier-Stokes equations if $(u,v,p)(x,y)$ is a solution, see e.g. \citet{budanur2015}. 
\par
Discrete symmetry groups, $\bb{G}$, are a finite set of operators, that transform the solution in a distinct way for each member of the set. An example is given by the group of reflections, $\bb{Z}$, where $(u,-v)(x,-y)$ is a solution to the underlying equations, if  $(u,v)(x,y)$ is a solution. This symmetry is present in e.g. the two dimensional flow around a cylinder. In the presence of multiple discrete symmetry groups, these can be unified into one by creating a new group $\bb{G'}$, that is composed of all unique combinations of the original symmetry groups. Note, that in general the action of two symmetry groups does not permute.
\par
These types of symmetries pose issues when performing model reduction on a discretized system, since approximately identical states, mainly differing in an application of a symmetry operation, are being encoded into vastly different regions of the latent space, see e.g. \citet{Page_2021}.
Additionally, the PCA basis for these types of problems is less interpretable due to the restrictions placed on the modes by the invariant solutions. In the following, a brief description of linear and nonlinear model reduction methods shall be given, followed by an excursion into the interpretability problems arising from symmetry copies.

\subsection{PCA and NLPCA}
An autoencoder (AE) is a parametric model that encodes a given input dataset $\mb{X}\in \mathbb{R}^{N_{\bb{\upxi}}\times N_t}$, dependent on spatial coordinates $\bb{\upxi}$ and time $t$ of size $N_{\bb{\upxi}}$ and $N_t$, into a latent space $\mb{A}\in \mathbb{R}^{N\times N_t}$, and decodes this latter description back to the input space, hence creating a reconstruction $\mb{\hat{X}}(\bb{\upxi},t)$. The latent space is designed to have a dimension $N$ that is lower than that of the input and output space, thus, achieving a compression of the data. This parametric model, where $E$ and $D$ denote the encoder and decoder, respectively, can be expressed as 
\begin{align}
        \mb{A} &= E(\mb{X})\\
        \mb{\hat{X}}&=D(\mb{A}).
\end{align}
The parameters of this model, i.e. the weights of the network, can be trained by backpropagating the $L_2$-error for each sample, $|\mb{\hat{X}}-\mb{X}|_2 $. 
If this parametric model is linear, i.e. $E$ and $D$ are matrices that are being multiplied with the input and the latent space, an analytical method to train this model is given by PCA, see \citep{BALDI1989} and \citep{plaut2018}.
These models, however, do not have to be linear and show larger compression when they are not. An introduction of how such a network can be constructed can be found in e.g. \citep{Milano2002} and \citep{murata2020}. For our investigations we constructed a a hybrid AE employing both convolutional layers as well as fully connected layers described in \cref{tab:app_kolm_2}.
\begin{table}[ht!]
\centering
\begin{tabular}{|l|l|l|l|}
\hline
\multicolumn{4}{|c|}{\textbf{Architecture}}                                                                                       \\ \hline
\multicolumn{2}{|c|}{Encoder}                    & \multicolumn{2}{c|}{Decoder}                     \\ \hline
\multicolumn{1}{|c|}{Layer}                         & \multicolumn{1}{|c|}{Data size}        & \multicolumn{1}{|c|}{Layer}                         & \multicolumn{1}{|c|}{Data size}        \\ \hline
Input                         & (Batch, 32, 32)    & Latent space                  & (Batch, N)        \\ \hline
Reshape                       & (Batch, 32, 32, 1)  & Fully connected (swish)       & (Batch, 2048)     \\ \hline
Convolution ((3, 3, 16),  swish) & (Batch, 32, 32, 16) & Fully connected (linear)      & (Batch, 4096)     \\ \hline
Average pooling (2, 2)         & (Batch, 16, 16, 16) & Reshape                       & (Batch, 8, 8, 64)   \\ \hline
Convolution ((3, 3, 32),  swish) & (Batch, 16, 16, 32) & Upsampling (2, 2)              & (Batch, 16, 16, 64) \\ \hline
Average pooling (2, 2)         & (Batch, 8, 8, 32)   & Convolution ((3, 3, 32),  swish) & (Batch, 16, 16, 32) \\ \hline
Convolution ((3, 3, 64),  swish) & (Batch, 8, 8, 64)   & Upsampling (2, 2)              & (Batch, 32, 32, 32) \\ \hline
Reshape                       & (Batch, 4096)     & Convolution ((3, 3, 16),  swish) & (Batch, 32, 32, 16) \\ \hline
Fully connected (swish)       & (Batch, 2048)     & Convolution ((3, 3, 1),  linear) & (Batch, 32, 32, 1)  \\ \hline
Fully connected (linear)      & (Batch, N)        & Reshape                       & (Batch, 32, 32)    \\ \hline
Latent space                  & (Batch, N)        & Output                        & (Batch, 32, 32)    \\ \hline
\end{tabular}
\\
\vspace{0.5cm}
\begin{tabular}{|l|l|l|l|}
\hline
\multicolumn{4}{|c|}{\textbf{Hyperparameters}}                                                                                       \\ \hline
\multicolumn{1}{|c|}{Parameter} & \multicolumn{1}{c|}{Value} & \multicolumn{1}{c|}{Parameter} & \multicolumn{1}{c|}{Value}  \\ \hline
Maximum epochs                  & 10                         & Early stopping delta           & $10^{-7}$                    \\ \hline
Learning rate                   & $10^{-3}$                     & Batch size                     & 500                         \\ \hline
Early stopping patience         & 5                          & Optimizer                      & Adam (Keras default values) \\ \hline
\end{tabular}
\caption{Autoencoding architecture and hyperparameters used to train the autoencoding architecture for NLPCA and I-NLPCA for the Kolmogorov flow dataset.}\label{tab:app_kolm_2}
\end{table}
The reader should note this is by no means the best such architecture that can be found for our problem. Rather, it was found that it performs reasonably well in compressing errors. The advantages of the symmetry-aware embedding do not depend on having constructed the best autoencoder.
\subsubsection{PCA and invariant solutions}
For the following discussion an ergodic turbulent flow with statistical stationarity and a equal appearance of the symmetry transformed states is assumed. While this is not a necessity for the arising problem of PCA modes being constricted in their shape due to invariant solutions, it is required for the exact way the shapes are constricted.
\par
If the continuous group of translations, $\bb{T}(s)$, is applicable to a specific coordinate direction, the PCA will return a Fourier basis in this direction, see e.g. \citet{holmes2012}. This property is often exploited when periodic boundaries are present, since one can perform a discrete Fourier transform (DFT) in the periodic direction prior to performing PCA and hence reduce the computational complexity of the operation, see e.g. \citet{Freund2009}. These modes are arguably less informative with regard to flow structures, since they only mirror the translational behavior which is already known due to the homogeneous direction.
\par
Given discrete symmetries a similar issue emerges. For the discrete reflectional symmetry, $\bb{Z}$, the PCA basis displays symmetric or antisymmetric shapes with respect to the axis of reflection, see \citet{SIROVICH1987}. Turbulence initially breaks the symmetry of the laminar base flow. After a sufficiently long time, however, these symmetries are statistically recovered, i.e., each solution appears in its original and reflected state, approximately equi-distributed. This property has again been exploited for reducing the computational requirements to calculate the PCA modes, see e.g. \citet{bourgeois2013}. In the case of reflectional symmetries, we can divide the flow into a symmetric and an antisymmetric part, each only half the size of the original domain, reducing the complexity of the eigenvalue problem. However, since modes are either symmetric or antisymmetric, it might be tempting to conclude that each sample is either symmetric or antisymmetric, when in fact the majority of samples are asymmetric.
\section{Symmetry-aware model reduction: s-PCA and s-NLPCA}
\begin{figure}[t]
  \centering
         \centering
         \includegraphics[width=\textwidth]{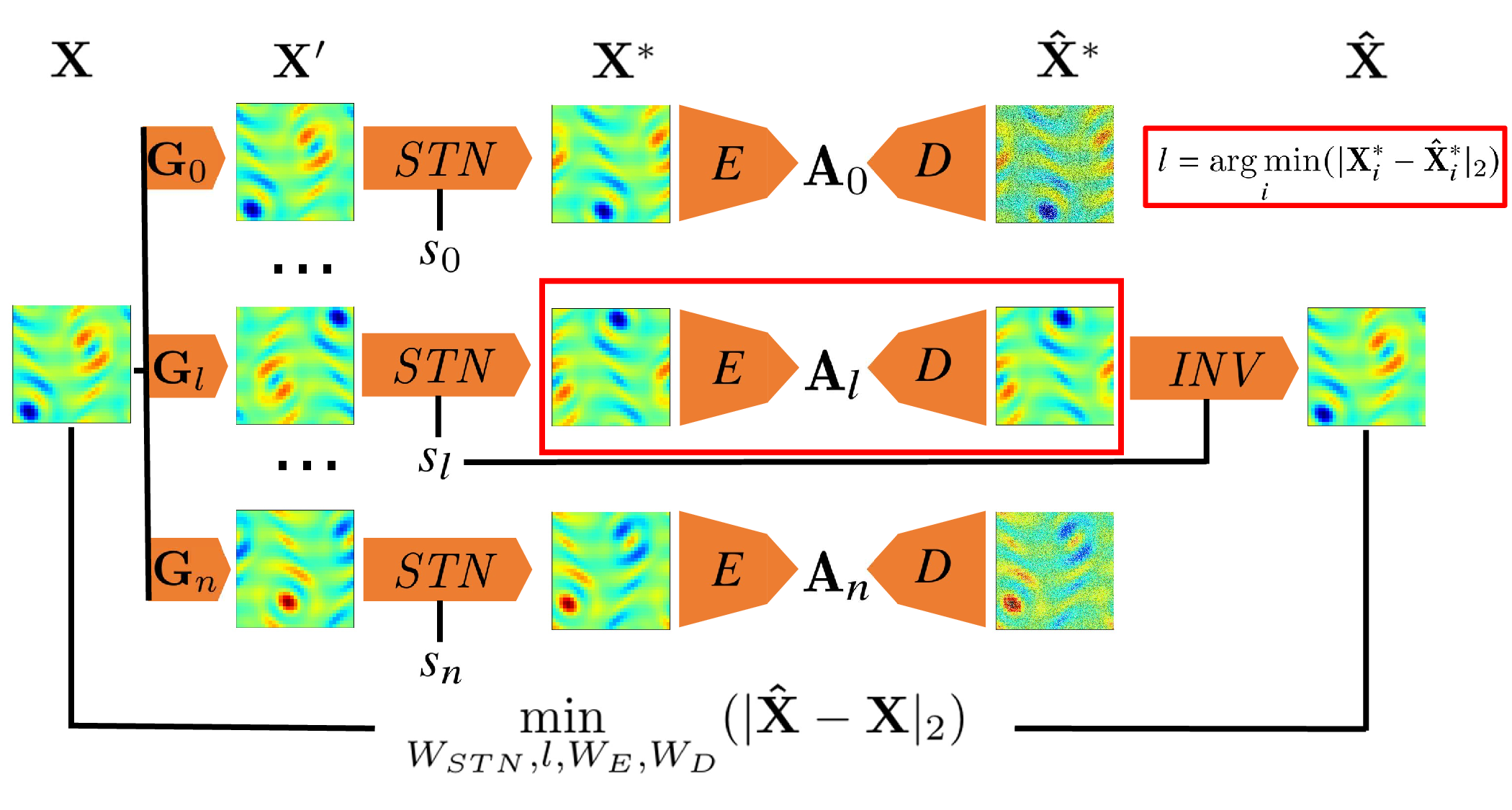}
\caption{Visualization of the symmetry-aware autoencoder architecture developed in this paper. A sample $\mb{X}$ is passed to the network. Each member of a set of predefined {discrete} transformations, $\mb{G}$, is applied to the input creating siamese branches ($\mb{X}\rightarrow \mb{X'}$). A secondary network, called an \textit{STN} \citep{jaderberg2015}, then extracts a continuous transformation parameter $s$ from each of the discretely transformed samples ($\mb{X'}\rightarrow s$) and manipulates these according to a predefined \textbf{continuous} transformation function, $\mb{F}(s)$($\mb{X'}\rightarrow \mb{X^*}$). The same autoencoder is then applied to all branches in parallel ($\mb{X^*}\rightarrow \mb{\hat{X}^*}$). The resulting reconstructions are then merged by selecting the branch, $l$, with the least $L_2$-error to its respective input ($\mb{\hat{X}^*}\rightarrow \mb{\hat{X}}^*_l$). The inverse continuous symmetry group $\smash{\mb{F}(s)^{-1}}$, as well as the inverse discrete transform $\mb{G}^{-1}_l$ are then used to retransform the reconstruction to the original input space (${INV}:\smash{\mb{\hat{X}^*}_l\rightarrow \mb{\hat{X}}}$). The trainable parameters of the network, i.e. the \textit{STN} and the autoencoder, are trained by backpropagating the $L_2$-error, $|\mb{{X}}^*_l-\mb{\hat{X}}^*_l|_2$.}
\label{fig:sAE}
\end{figure}

To remove symmetry copies from the solution set in an automated manner, a neural network embedding has been designed and is shown in \cref{fig:sAE}. Please see {\color{blue}\href{https://github.com/simonkneer/Symmetry-Aware-Autoencoding}{https://github.com/simonkneer/Symmetry-Aware-Autoencoding}} for Keras implementations of this network for different fluid problems.
The schematic shows how our proposed architecture deals with continuous and discrete symmetry groups simultaneously. When the batch of samples $\mb{X}$ is first passed to the network, the states are manipulated according to the discrete symmetry group that is hardcoded into the network, according to the discrete symmetries in the system of equations,
{\color{black}
\begin{equation}
    [\mb{{X}'}_0\dots \mb{{X}'}_i \dots \mb{{X}'}_n]=[\mb{G}_0\mb{{X}}\dots \mb{G}_i\mb{{X}}\dots \mb{G}_n\mb{{X}}].
\end{equation}}
A network with split branches each containing the same subnetworks is known as a Siamese network. We note that the weights of the networks are identical across each of these parallel branches. In the case of reflectional symmetries, two branches are created, where $\mb{G}_0$ is the identity and $\mb{G}_1$ is the reflection operation $(u,v)(x,y)\rightarrow (u,-v)(x,-y)$.
\par
A subnetwork, called a Spatial Transformer Network (\textit{STN}) \citep{jaderberg2015}, evaluates each of these branches. This subnetwork is in itself an encoder that transforms each of the symmetry copies to a set of transformation parameters $\mb{p}_i$: 
\begin{equation}
        \mb{p}_i = STN(\mb{X'}_i).\label{eq:STN}
\end{equation}
As stated this $STN$ is in itself just an encoder and as such can be realized with different types of neural networks. In our case this encoder simplifies to just a projection as described below. {\color{black}The samples $\mb{X'}_i$ are then modified according to the prescribed continuous transformation function $\mb{F}(\mb{p}_i)$,
\begin{equation}
    \mb{X}^*_i=\mb{F}(\mb{p}_i)\mb{X'}_i\quad \forall i.
\end{equation}
}
For our special case of Kolmogorov flow, the $STN$ only identifies a shift in the $x$-direction, which is a linear transform. Hence, the extraction of the transformation parameter can be achieved with a linear network, i.e. the projection of the states onto the weights of the network. Additionally, it is apparent that the shift is a cyclic variable due to the periodic boundaries. Thus, the \textit{STN} is set to predict two output parameters, $s_{1,i}$ and $s_{2,i}$,
\begin{align}
    s_{1,i} &= \mb{W}_{s,1}\mb{X'}_i\\
    s_{2,i} &= \mb{W}_{s,2}\mb{X'}_i    
\end{align}
which are then merged following
\begin{equation}
    s_i=\text{arg}(s_{1,i}+is_{2,i})\frac{{L}_{{x}}}{2\pi}.\label{eq:shift_combine}
\end{equation}
If one lets this network train without further intervention it will naturally converge to a state where the weights predicting $s_2$ are approximately those predicting $s_1$ shifted by a quarter domain, as described for the method of slices using the first Fourier slice \cite{budanur2015exact}. Thus the $STN$ has naturally assumed the form of the method of slices. In practise it then makes sense to constrain the weight of $s_2$ to that condition. At this point it also becomes clear why we have to apply the $STN$ to all the siamese branches. Naively one might assume that we can apply the $STN$ prior to splitting the network into siamese branches. However, the action of the discrete transforms will always kick the sample out of the slice in some way, since there is no slicing template that is invariant under both rotations $\bb{P}$ and the shift and reflect symmetry  $\bb{S}$.
The autoencoder is then applied to each of discretely and continuously transformed samples,
\begin{align}
    [\mb{A}_0\dots \mb{A}_l \dots \mb{A}_n]&=[E(\mb{X}^*_0)\dots E(\mb{X}^*_i) \dots E(\mb{X}^*_n)]\\
    [\mb{\hat{X}}^*_0\dots \mb{\hat{X}}^*_i \dots \mb{\hat{X}}^*_n]&=[D(\mb{A}_0)\dots D(\mb{A}_i) \dots E(\mb{A}_n)].
\end{align}
\par
As a next step, we identify the branch, $l$, that has the lowest reconstruction error with respect to its input to the autoencoder, i.e.
\begin{equation}
    l=\argminA_i (|\mb{X}^*_i-\mb{\hat{X}}^*_i|_2).
\end{equation}
Lastly, the reconstruction with the lowest error is retransformed into the input space. In other words, we apply the inverse of the continuous symmetry group for the parameter $\mb{s}_l$
\begin{equation}
    \mb{\hat{X}'_l}=\mb{F}^{-1}(\mb{s}_l)[\mb{\hat{X}^*_l}].
\end{equation}
The inverse of the discrete symmetry operation that was used for the creation of the chosen branch, {\color{black}
\begin{equation}
    \mb{\hat{X}}=\mb{G}^{-1}_{l}\mb{\hat{X}'}_l.
\end{equation}}
is then applied.
As a result, we achieved a reconstruction of the original sample in the input space with all invariant solutions encoded into the same latent space. For a full description of the input the shift parameter $s_l$ is now part of the latent space as well. Note that for training the weights we calculate the error $|\mb{X}^*_l-\mb{\hat{X}}^*_l|_2$ and backpropagate it only through the chosen branch and the \textit{STN}.
\par
When comparing the computational effort of this model to that of a conventional AE design we do not notice a significant increase in cost. While the network calculates the forward pass of each sample in parallel, which is comparable to the cost of a matrix multiplication, the more costly backpropagation is performed on only one branch (with the minimal error). As a result, the parallel branches of the Siamese network do not substantially increase the cost. The \textit{STN}, however, requires a small additional training effort, equivalent to an encoder for a two-dimensional latent space.\\
In practice,however, it is convenient to perform this optimization with a linear autoencoder network with only one mode. Once we have found the optimal slicing template as well as the fundamental domain, we perform other model reduction methods on $\mb{X}^*_l$, where $l$ is the discrete branch with the lowest reconstruction error.

\section{s-PCA and s-NLPCA for Kolmogorov flow}\label{sec:kolmogorov}
\begin{figure}[t!]
\centering
\begin{subfigure}{0.45\textwidth}
         \centering
         \includegraphics[width=\textwidth]{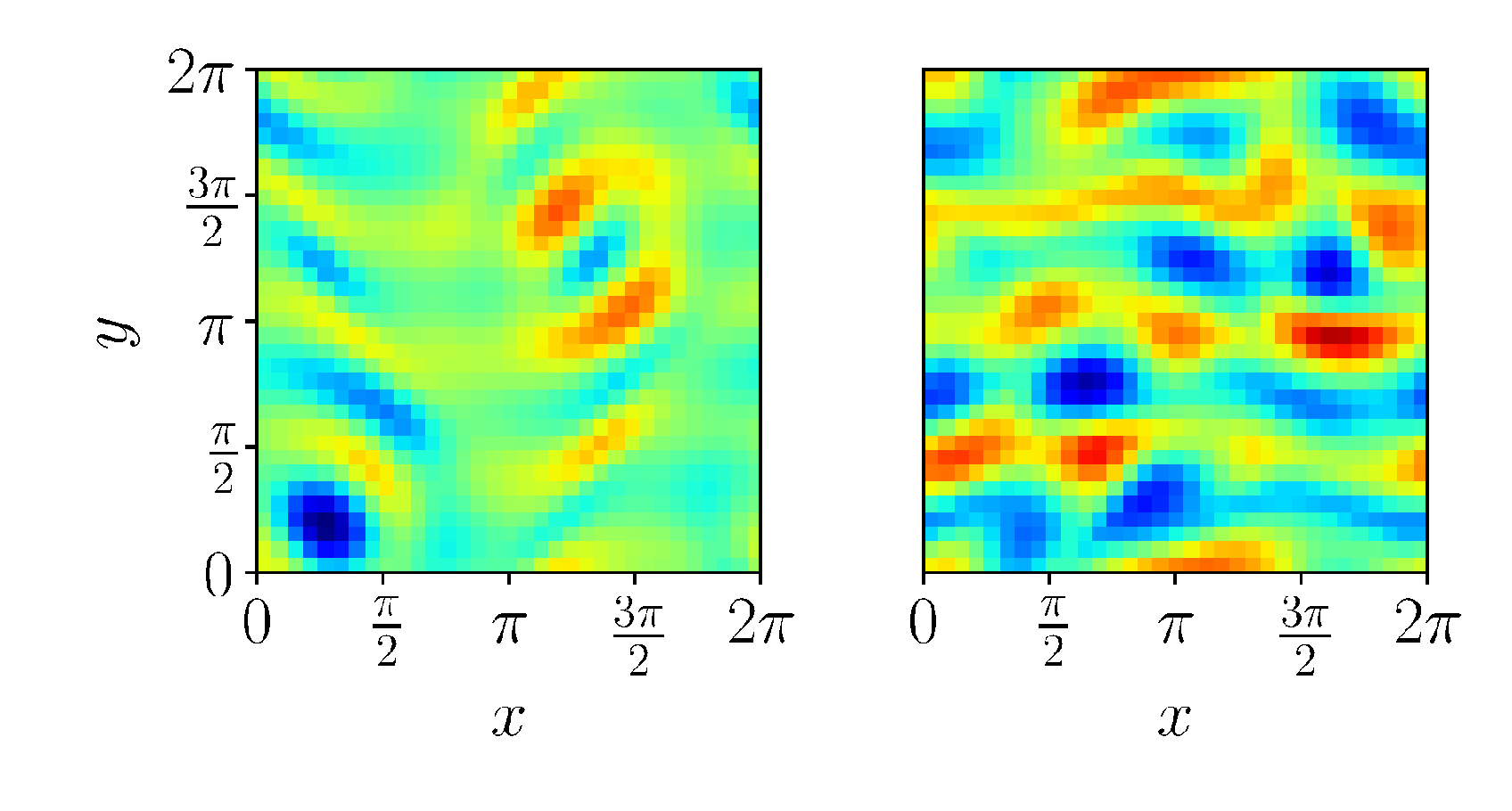}
\end{subfigure}
\begin{subfigure}{0.54\textwidth}
         \centering
         \includegraphics[width=\textwidth]{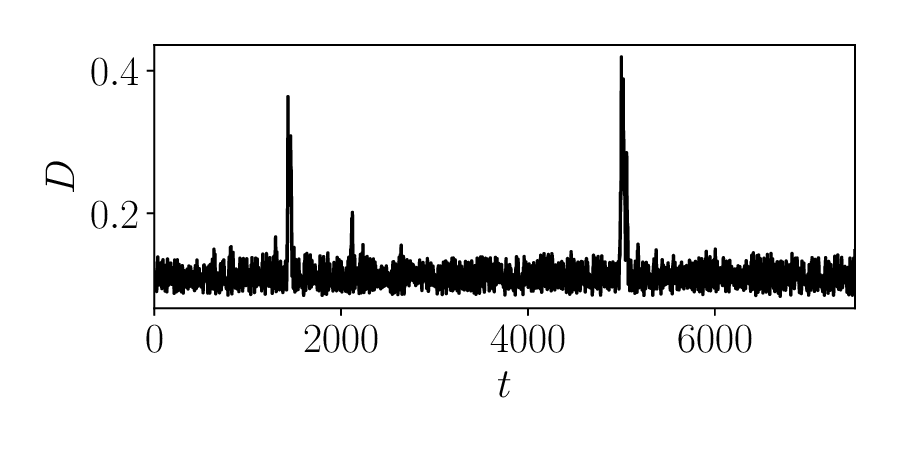}
\end{subfigure}
\caption{Snapshots of vorticity, ${\omega}$, for Kolmogorov flow with $n=4$ at Re$=35$ at two different times, exhibiting low (left) and high (middle) dissipation, as well as an exemplary time series of dissipation $D$ (right).}
\label{fig:KOLMO}
\end{figure}

We shall now investigate the effectiveness of our proposed embedding by applying it to Kolmogorov flow, an example that involves both continuous and discrete symmetry groups. To generate data, we used the code by \citet{Wan_2018}, which simulates Kolmogorov flow, arising from the sinusoidally forced Navier-Stokes equations with periodic boundary conditions in both $x$ and $y$, first proposed by \citet{MESHALKIN1961}. We employ the cross plane vorticity formulation 
\begin{align}
    \frac{\partial \omega}{\partial t} = -\mb{u}\cdot\nabla \omega+\frac{1}{\text{Re}}\nabla^2\omega-n\cos(ny),
\end{align}
where $\omega$ is the vorticity, $\mb{u}$ the velocity vector, Re the Reynolds number and $n\cos(ny)$ the sinusoidal forcing term.
This set of equations is often used to mimic turbulence in two dimensions. We choose a domain size of $32\times 32$, a forcing wavenumber of $n=4$ in the $y$-direction and Re$=35$, which leads to a chaotic regime (Re$_{crit}=33.6$ \citep{chandler2013}). This can be seen in the bursting behavior of the dissipation, $D$, visualized in \cref{fig:KOLMO}. We collect 300,000 samples by integrating a random initial condition with $\Delta t_{int}=0.005$ and recording vorticity fields in intervals $\Delta t_{rec}=0.1$. {\color{black}Due to the large size of the dataset and the number of discrete symmetries, data manipulation based on symmetry transformations, as proposed by \citep{holmes2012}, was not performed.}
Vorticity snapshots, for a low- and high-dissipation event, as well as temporal evolution of the dissipation, are shown in \cref{fig:KOLMO}.
\par
Due to the periodic boundary conditions and the wavenumber of $n=4$ for the forcing in $y$, multiple symmetry groups exist:
the group of continuous translations in $x$, i.e. $\bb{T}(s)$: $\omega(x,y)\rightarrow\omega(x+s,y)$, the 8-cyclic shift and reflect symmetry $\bb{S}$: $\omega(x,y)\rightarrow\omega(-x,y+\pi/4)$, and the 2-cyclic rotation symmetry, $\bb{P}$: $\omega(x,y)\rightarrow\omega(-x,-y)$, see e.g. \citet{chandler2013}. In what follows, we combine the two discrete symmetry groups into a group $\bb{G}=[\bb{P}_0\bb{S}_0,\dots,\bb{P}_0\bb{S}_7,\bb{P}_1\bb{S}_0,\dots, \bb{P}_1\bb{S}_7]$, so that we can construct a network as shown in \cref{fig:sAE}.
\par
\begin{figure}[t!]
  \centering
         \includegraphics[width=0.45\textwidth]{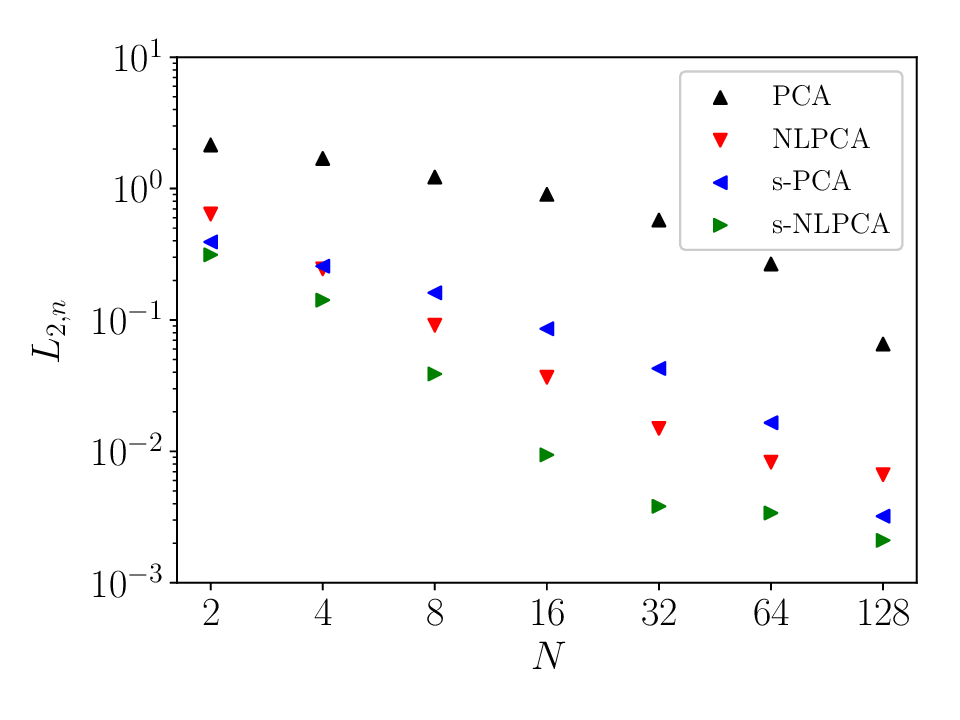}
\caption{$L_2$ reconstruction error for PCA, NLPCA, s-PCA and s-NLPCA for Kolmogorov flow versus number of modes used for the reconstruction.}
\label{fig:kolmo_l2}
\end{figure}

The $L_2$ reconstruction errors, $|\mb{X}-\mb{{X}}|_2$, for all considered methods, i.e. PCA, s-PCA, NLPCA and s-NLPCA, are visualized in \cref{fig:kolmo_l2}.
This figure shows that all alternative methods outperform PCA by a wide margin. While in PCA the error starts to decrease in a meaningful way for $N>32$, s-PCA exhibits a larger decrease from the start and shows a decrease in the slope for higher $N$, which is absent for PCA. The error for NLPCA is initially higher than for s-PCA, but crosses below the latter at $N=2$, to then decrease with a similar slope. It then tapers off at $N=64$. At $N=32$ the error for s-NLPCA reaches a quasi minimum. While strictly speaking the error at $N=128$ is lower than that at $N=32$, it is not by a large amount. Hence, it seems fair to assume that the most significant features of the flow are captured at $N=32$. We believe a further refinement of our network architecture could lead to the error not decreasing past $N=32$.
\par
\begin{figure}[t!]
\centering
\begin{subfigure}{0.78\textwidth}
         \centering
         \includegraphics[width=\textwidth]{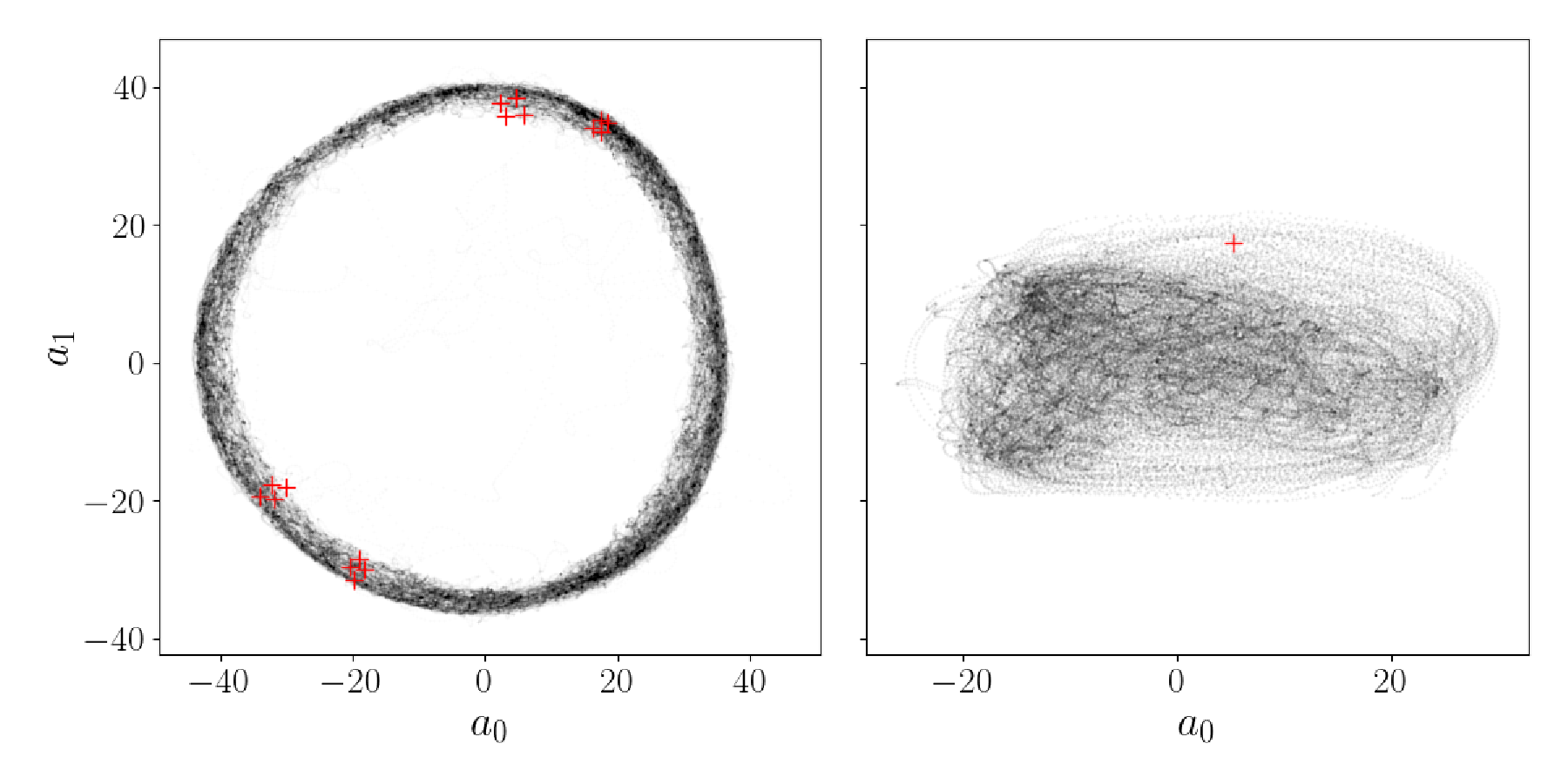}
\end{subfigure}
\caption{PCA amplitudes of mode 1 against those of mode 2 for the input space (left) and the symmetry reduced space (right). Red crosses are discrete symmetry copies of a randomly chosen state.}
\label{fig:trajectories}
\end{figure}

In order to illustrate why the symmetry reduction leads to a decrease in error for this system we show the PCA amplitudes of mode 1 against those of mode 2 for both the input space and the symmetry reduced space in \cref{fig:trajectories}.
It is easy to see why the reconstruction error for the symmetry reduced methods is lower. While the inclusion of a translational symmetry only increases the dimensionality of the system by one, this translational behaviour is hard to be captured by only one PCA mode or a latent variable. In fact this behavior heavily dominates the first few PCA components, reflected in the circular trajectory of the PCA amplitudes in the unreduced space. While discrete symmetries do not increase the dimensionality of our system they do reshape the manifold on which the flow resides. Since less complicated structures are more easily captured by data driven reduction methods we perceive a reduction in error. In \cref{fig:trajectories} we also see how the symmetry copies collapse into one point when the full reduction methods are applied. Hence, the manifold is less complicated and has indeed no duplicities.
\par
\begin{figure}[t!]
     \centering
     \begin{subfigure}{1\textwidth}
         \centering
         \includegraphics[width=\textwidth]{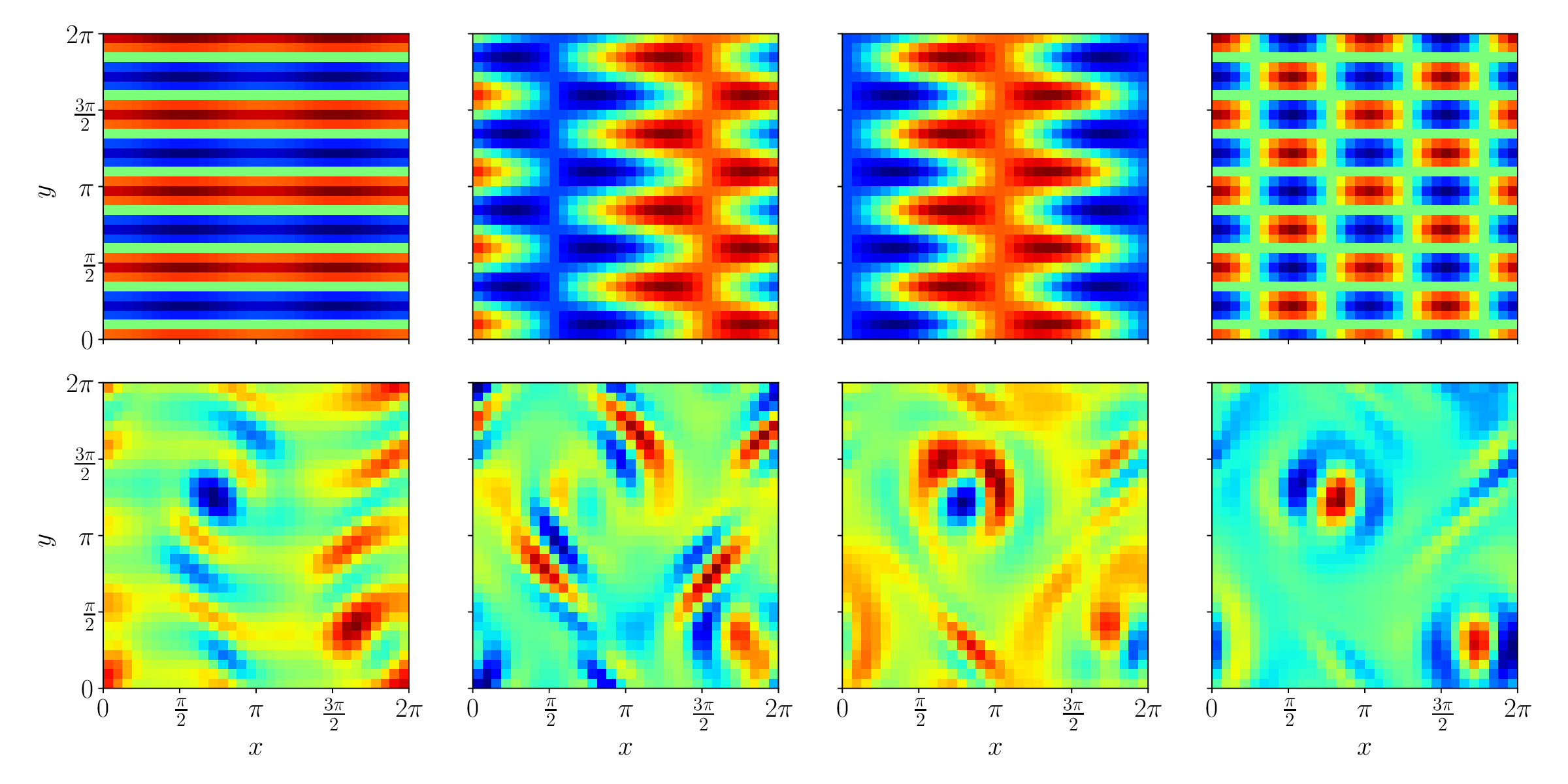}
     \end{subfigure}
     \caption{Vorticity modes for Kolmogorov flow using PCA (top) and s-PCA (bottom) for $i\in\{Mean,1,2,3\}.$}
     \label{fig:kolmo_modes}
\end{figure}

In \cref{fig:kolmo_modes} we show the temporal mean of the input vorticity fields and the symmetry reduced ones, alongside the first 3 PCA components for each dataset.
We observe that the mean of the vorticity as well as all present PCA modes exhibit a wavenumber $n=4$ pattern in the $y$ direction stemming from the forcing wavenumber As expected, the PCA modes assume a Fourier-like shape in the $x$-direction. As discussed above and visualized in the trajectory of their projections in \cref{fig:trajectories}, modes 1 and 2 form a mode pair that together reflects a translational dynamic. Mode 3 is merely a higher harmonic of the translation in $x$ as well as a reflection of the shift and reflect symmetry. Calling on \cite{holmes2012} these shapes adhere exactly to the condition, that the space spanned by the modes be invariant under the symmetry operations of the system. It is obvious that this basis is not very representative of the low-dissipation snapshot in \cref{fig:KOLMO}, in whose vicinity the solution resides for a majority of time. When moving to s-PCA, a different behavior emerges. The mean now represents a distorted version of the equilibrium state at this Reynolds number; see e.g. \citet{platt1991} for this equilibrium state. This mean in addition to the modes visualized here are a linear modal representation of a relative periodic orbit embedded in the turbulent flow similar to the one reported in the supplementary material of \cite{chandler2013}. Here we observe a counter rotating vortex pair that is advected. It is reasonable to assume that this RPO is being shadowed by the turbulence for a long time, since it is being represented in the primary PCA modes.
\par
\begin{figure}[t!]
\centering
\begin{subfigure}{0.8\textwidth}
         \centering
         \includegraphics[width=\textwidth]{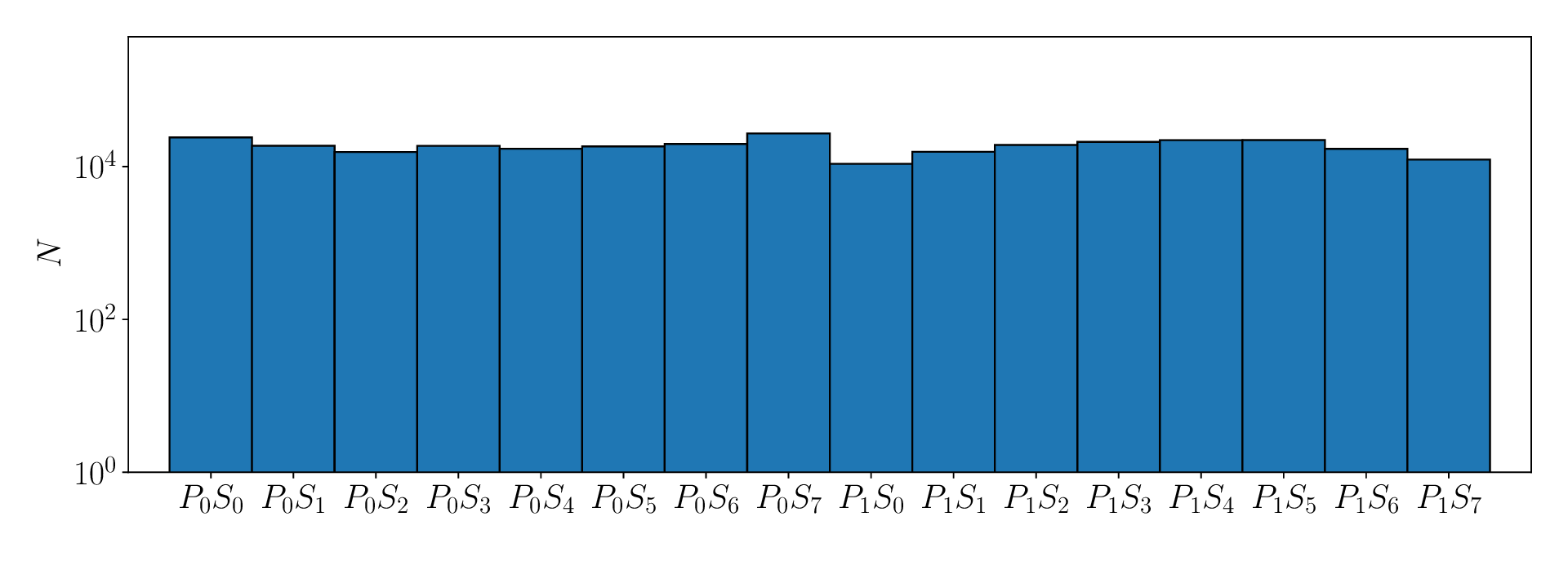}
\end{subfigure}
\caption{Number of times each branch was chosen as the one generating the minimum error, $l$.}
\label{fig:kolmo_shift}
\end{figure}

In \cref{fig:kolmo_shift} we show the number of times each branch was chosen as the one with the lowest reconstruction error, $l$.
For the discrete-symmetry-reducing step there exist 16 symmetry transformations, $\bb{G}=[\bb{P}_0\bb{S}_0,\dots,\bb{P}_0\bb{S}_7,\bb{P}_1\bb{S}_0,\dots, \bb{P}_1\bb{S}_7]$, resulting in 16 Siamese branches for the autoencoder, which allows us to compute the number of times, $N$, each branch was chosen as the minimum-error option. This provides insight into the  statistical convergence of our dataset in terms of the discrete invariant transformations. We observe that each of the branches was chosen in roughly equal amounts. Hence, our choice not to symmetrize the data as suggested by \cite{holmes2012} appears reasonable. At the same time it uncovers the necessity of symmetry reduction in even this weakly turbulent flow. After only 30,000 fluid timesteps the distribution of states in different symmetry orientations has almost reached an equal distribution. As can be seen by the PCA components in \cref{fig:kolmo_modes} this leads to spaces spanned by the PCA modes, that are invariant with respect to the discrete symmetries of our system and poor insight into the physics.
\par
\begin{figure}[t!]
\centering
\begin{subfigure}{0.45\textwidth}
         \centering
         \includegraphics[width=\textwidth]{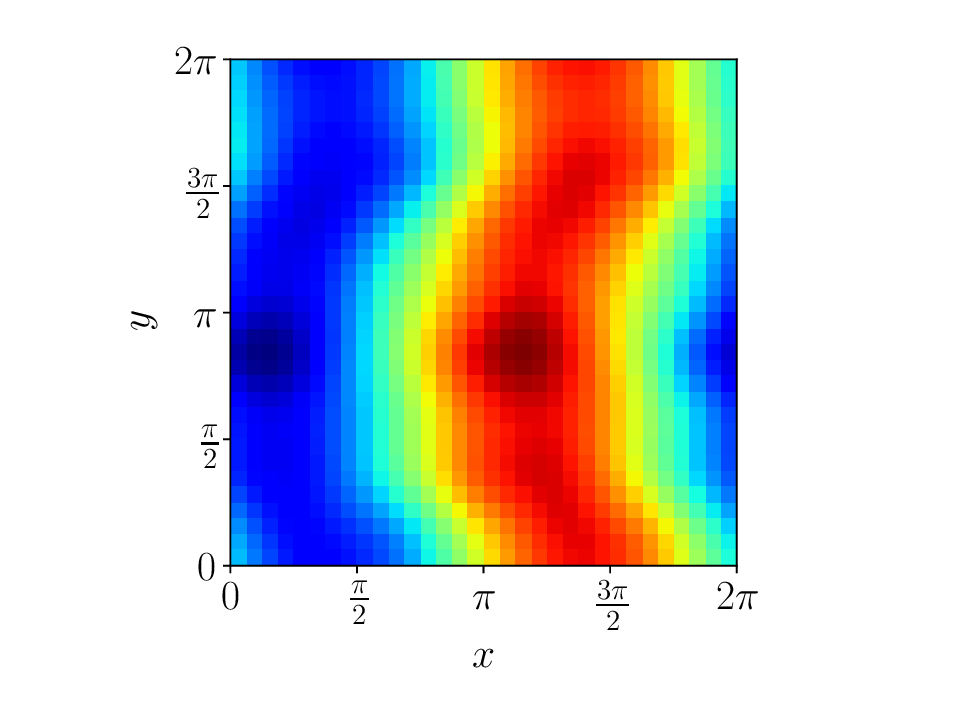}
\end{subfigure}
\begin{subfigure}{0.45\textwidth}
         \centering
         \includegraphics[width=\textwidth]{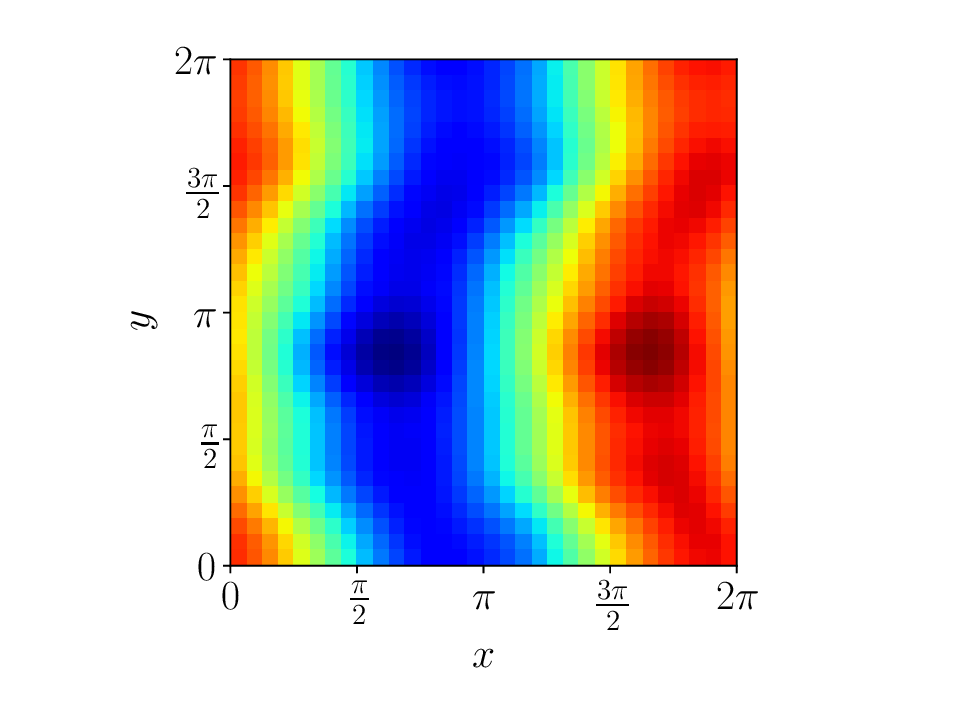}
\end{subfigure}
\caption{Slice templates $\mb{W}_{s,1}$ (left) and $\mb{W}_{s,2}$ (right) learned from data.}
\label{fig:kolmo_weights}
\end{figure}

We anticipate the slicing templates, i.e. the weights of the $STN$, to converge to trigonometric functions in the translated direction, even though a prediction on the shape of the weights in the inhomogeneous direction $y$ cannot be made. According to \citet{marensi2021}, a dependence of the transformation parameter on the non-translated, inhomogeneous directions may be required, if discontinuities of the predicted shift are to be avoided. Investigating the weights reveals that such a dependence can theoretically be learned, without any further user input. The learned slice templates are compiled in \cref{fig:kolmo_weights}.
As expected, the slicing templates take on a shape similar to $\cos x$ and $\sin x$. However, we do notice a modulation of the weights in the $y$-direction. To aid convergence, we prescribed that $\mb{W}_{s,2}=\bb{T}(L_x/4)\mb{W}_{s,2}$.Prior to this, however, we did not constrain the network to this condition, which lead to it approximately discovering it by itself.
\par
\begin{figure}[t!]
\centering
\begin{subfigure}{0.54\textwidth}
         \centering
         \includegraphics[width=\textwidth]{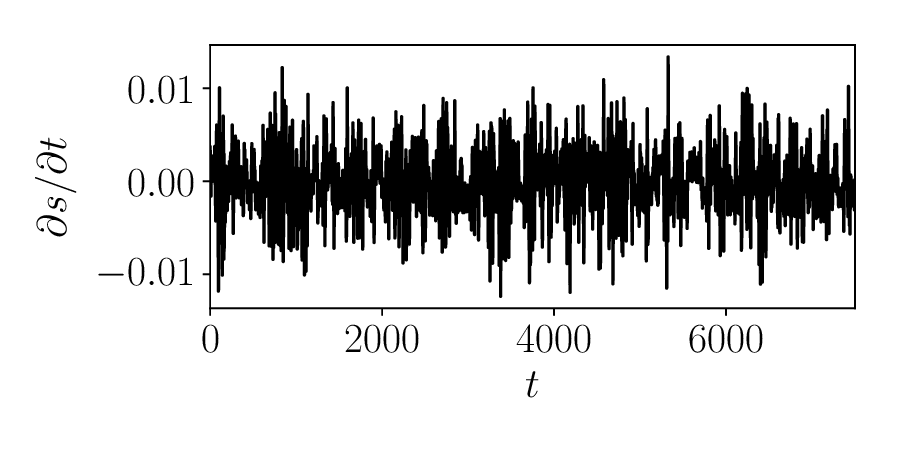}
\end{subfigure}
\caption{Exemplary time series of the temporal gradient of the shift from the $STN$.}
\label{fig:kolmo_s_grad}
\end{figure}

To investigate the effectiveness of this modulation in the inhomogeneous direction, we will investigate the gradient of the shift $\frac{\partial s}{\partial t}$. A well designed or learned slicing template should not result in large gradients of the shift. In order to show this, not the shifts $s$ from the network itself were extracted, as these are naturally discontinuous in time due to the switching between branches, but rather the original input data was fed into the trained $STN$. The exemplary resulting temporal gradient of the shift is shown in \cref{fig:kolmo_s_grad}.
It is apparent that the gradient of the shift does not explode at any point in our timeseries. Thus it appears as though the $STN$ was able to find a suitable slice template for this dataset. It might be beneficial to explore this behavior further on a case where first Fourier mode slicing fails more often.
\section{Conclusion}
In this work we present a novel, symmetry-aware machine learning approach for symmetry and model reduction of physical problems. For both linear and nonlinear methods our symmetry-aware embedding was able to reduce the $L_2$ reconstruction error by large margins when compared to more conventional methods. Hence, a comparable accuracy of reconstructions could be achieved with far fewer components, which makes each individual one more relevant and worth investigating. Our method was tested on Kolmogorov flow, which is characterized by a complex scenario: a combination of the translational symmetry $\bb{T}(s)$ in $x$, along with two discrete symmetry groups, namely a rotation through $\pi$, $\bb{P}: \omega (x,y)\rightarrow \omega(-x,-y)$ and a shift-and-reflect symmetry, $\bb{S}:\omega(x,y)\rightarrow\omega(-x,y+\pi/4)$.
\par
Our method is based on transforming our input dataset into a new space that is invariant with respect to symmetry transformations, and performing model reduction in that space. 
To remove the discrete symmetry copies from the dataset a Siamese network is formed. After the removal of the continuous symmetries and the autoencoding process, we prune our network to one single branch by selecting the branch with the lowest reconstruction error relative to its respective input. Thus, when a sample reappears in a different discrete symmetry state, it is automatically reoriented, since the AE has been previously conditioned to compress a deviating orientation. This is in effect an autonomous way of transforming our data in to a fundamental domain, hence, not introducing human bias.
\par
Continuous symmetries are being removed using an \textit{STN}-like method. Here, a subnetwork encodes the state to a continuous transformation parameter $s$, the translational shift, which is then applied through a prescribed transformation function. Since the transformation in the case of Kolmogorov flow is merely a translation, hence linear, the \textit{STN} does not necessarily need to be nonlinear in order to discover the proper shifts for this dataset. A linear, which the $STN$ is in our case, is merely a projection of the state onto some weights. Consequently, the \textit{STN} is merely an optimization for an optimal slicing template. These templates resemble trigonometric functions (distinguished by a random phase shift) in the alignment direction. Furthermore, a modulation in the inhomogeneous direction, which does not change the alignment of the samples, can be used in cases such as turbulent channel flow, see e.g. \citet{marensi2021}, to avoid {\color{black}infinite shifts due to a vanishing Fourier amplitude for the first complex coefficient \citep{budanur2015}}. In a conventional application of the method of slicing this modulation has to be user-supplied and carefully tailored, whereas in our network it is learned as part of the network training.
\par
Within our symmetry-aware framework, both linear and nonlinear methods can be embedded. Applying a linear AE, we arrive at a method termed symmetry-aware linear PCA (s-PCA). Here, we do not achieve compression rates typical of nonlinear methods, but we gain insight into the dynamics of the system, as the modal structures are both extractable and interpretable.
Removing all symmetry groups for Kolmogorov flow changes the modes from an approximate Fourier basis, representative of the boundary conditions and the forcing wavenumber, to more interpretable structures. As a result, the temporal mean closely approximates the equilibrium solution, while higher modes recover the dynamics of a relative periodic orbit consisting of a vortex pair in the flow.
\par 
Using nonlinear methods leads to the method of symmetry-aware nonlinear PCA (s-NLPCA). The main gain here is a substantial increase in compression over conventional nonlinear PCA (NLPCA). Using 32 s-NLPCA modes for the dataset generated by Kolmogorov flow yields a quasi minimal reconstruction error that does not substantially decrease by increasing the amount of latent variables.
\par
Whether the learned slicing templates of our \textit{STN} weights in the non-aligned direction, can indeed circumvent the issue of infinite shifts for vanishing first Fourier coefficients remains to be investigated in a future effort. In addition, s-NLPCA, with its high compression rate, should provide sufficient compression in order to properly cluster data in the latent space and thus get a characterization of different states.

\bibliographystyle{jfm}
\bibliography{jfm}

\begin{thebibliography}{24}
\expandafter\ifx\csname natexlab\endcsname\relax\def\natexlab#1{#1}\fi
\def\au#1{#1} \def\ed#1{#1} \def\yr#1{#1}\def\at#1{#1}\def\jt#1{\textit{#1}}
  \def\bt#1{#1}\def\bvol#1{\textbf{#1}} \def\vol#1{#1} \def\pg#1{#1}
  \def\publ#1{#1}\def\arxiv#1{#1}\def\org#1{#1}\def\st#1{\textit{#1}}

\bibitem[Baldi \& Hornik(1989)]{BALDI1989}
{\sc \au{Baldi, P.} \& \au{Hornik, K.}} \yr{1989}  \at{Neural networks and
  principal component analysis: {L}earning from examples without local minima}.
   \jt{Neural Networks}  \bvol{2}~(1),  \pg{53--58}.

\bibitem[Bourgeois {\em et~al.\/}(2013)Bourgeois, Noack \&
  Martinuzzi]{bourgeois2013}
{\sc \au{Bourgeois, J.~A.}, \au{Noack, B.~R.} \& \au{Martinuzzi, R.~J.}}
  \yr{2013}  \at{Generalized phase average with applications to sensor-based
  flow estimation of the wall-mounted square cylinder wake}.  \jt{Journal of
  Fluid Mechanics}  \bvol{736},  \pg{316–350}.

\bibitem[Budanur(2015)]{budanur2015exact}
{\sc \au{Budanur, N.~B.}} \yr{2015}  \at{Exact coherent structures in
  spatiotemporal chaos: from qualitative description to quantitative
  predictions}. PhD thesis, Georgia Institute of Technology.

\bibitem[Budanur {\em et~al.\/}(2015)Budanur, Cvitanovi\ifmmode~\acute{c}\else
  \'{c}\fi{} \& Siminos]{budanur2015}
{\sc \au{Budanur, N.~B.}, \au{Cvitanovi\ifmmode~\acute{c}\else \'{c}\fi{},
  P.~Davidchack, R.~L.} \& \au{Siminos, E.}} \yr{2015}  \at{Reduction of
  {SO}(2) symmetry for spatially extended dynamical systems}.  \jt{Phys. Rev.
  Lett.}  \bvol{114},  \pg{084102}.

\bibitem[Budanur \& Cvitanovi{\'{c}}(2016)]{Budanur2016}
{\sc \au{Budanur, Nazmi~Burak} \& \au{Cvitanovi{\'{c}}, Predrag}} \yr{2016}
  \at{Unstable manifolds of relative periodic orbits in the symmetry-reduced
  state space of the kuramoto{\textendash}sivashinsky system}.  \jt{Journal of
  Statistical Physics}  \bvol{167}~(3-4),  \pg{636--655}.

\bibitem[Chandler \& Kerswell(2013)]{chandler2013}
{\sc \au{Chandler, {G. J.}} \& \au{Kerswell, {R. R.}}} \yr{2013}  \at{Invariant
  recurrent solutions embedded in a turbulent two-dimensional {K}olmogorov
  flow}.  \jt{Journal of Fluid Mechanics}  \bvol{722},  \pg{554--595}.

\bibitem[Freund \& Colonius(2009)]{Freund2009}
{\sc \au{Freund, J.~B.} \& \au{Colonius, T.}} \yr{2009}  \at{Turbulence and
  sound-field {POD} analysis of a turbulent jet}.  \jt{International Journal of
  Aeroacoustics}  \bvol{8}~(4),  \pg{337--354}.

\bibitem[Fukami {\em et~al.\/}(2020)Fukami, Nakamura \& Fukagata]{fukami2020}
{\sc \au{Fukami, K.}, \au{Nakamura, T.} \& \au{Fukagata, K.}} \yr{2020}
  \at{Convolutional neural network based hierarchical autoencoder for nonlinear
  mode decomposition of fluid field data}.  \jt{Physics of Fluids}
  \bvol{32}~(9),  \pg{095110}.

\bibitem[Holmes {\em et~al.\/}(2012)Holmes, Lumley, Berkooz \&
  Rowley]{holmes2012}
{\sc \au{Holmes, P.}, \au{Lumley, J.~L.}, \au{Berkooz, G.} \& \au{Rowley,
  C.~W.}} \yr{2012} {\em Turbulence, Coherent Structures, Dynamical Systems and
  Symmetry\/}, 2nd edn.  \publ{Cambridge University Press}.

\bibitem[Jaderberg {\em et~al.\/}(2015)Jaderberg, Simonyan, Zisserman \&
  Kavukcuoglu]{jaderberg2015}
{\sc \au{Jaderberg, M.}, \au{Simonyan, S.}, \au{Zisserman, A.} \&
  \au{Kavukcuoglu, K.}} \yr{2015} {\em Spatial {T}ransformer {N}etworks\/}.

\bibitem[Linot \& Graham(2020)]{Linot2020}
{\sc \au{Linot, A.~J.} \& \au{Graham, M.~D.}} \yr{2020}  \at{Deep learning to
  discover and predict dynamics on an inertial manifold}.  \jt{Physical Review
  E}  \bvol{101}~(6).

\bibitem[Lumley(1967)]{lumley1967}
{\sc \au{Lumley, J.~L.}} \yr{1967}  \at{The structure of inhomogeneous
  turbulent flows}.  \jt{Atmospheric Turbulence and Radio Wave Propagation} .

\bibitem[Marensi {\em et~al.\/}(2021)Marensi, Yalnız, Hof \&
  Budanur]{marensi2021}
{\sc \au{Marensi, Elena}, \au{Yalnız, G\"okhan}, \au{Hof, Bj\"orn} \&
  \au{Budanur, Nazmi~Burak}} \yr{2021} Symmetry-reduced dynamic mode
  decomposition of near-wall turbulence.

\bibitem[Mehr {\em et~al.\/}(2018)Mehr, Lieutier, Bermudez, Guitteny, Thome \&
  Cord]{mehr2018}
{\sc \au{Mehr, \'{E}.}, \au{Lieutier, A.}, \au{Bermudez, F.~S.}, \au{Guitteny,
  V.}, \au{Thome, N.} \& \au{Cord, N.}} \yr{2018} Manifold learning in quotient
  spaces.  \bt{In {\em CVPR\/}},  \pg{pp. 9165--9174}.

\bibitem[Mendible {\em et~al.\/}(2020)Mendible, Brunton, Aravkin, Lowrie \&
  Kutz]{mendible2020}
{\sc \au{Mendible, A.}, \au{Brunton, S.~L.}, \au{Aravkin, A.~Y.}, \au{Lowrie,
  W.} \& \au{Kutz, J.~N.}} \yr{2020}  \at{Dimensionality reduction and
  reduced-order modeling for traveling wave physics}.  \jt{Theoretical and
  Computational Fluid Dynamics}  \bvol{34}~(4),  \pg{385–400}.

\bibitem[Meshalkin \& Sinai(1961)]{MESHALKIN1961}
{\sc \au{Meshalkin, L.~D.} \& \au{Sinai, Ia.~G.}} \yr{1961}  \at{Investigation
  of the stability of a stationary solution of a system of equations for the
  plane movement of an incompressible viscous liquid}.  \jt{Journal of Applied
  Mathematics and Mechanics}  \bvol{25}~(6),  \pg{1700--1705}.

\bibitem[Milano \& Koumoutsakos(2002)]{Milano2002}
{\sc \au{Milano, M.} \& \au{Koumoutsakos, P.}} \yr{2002}  \at{Neural network
  modeling for near wall turbulent flow}.  \jt{Journal of Computational
  Physics}  \bvol{182},  \pg{1--26}.

\bibitem[Murata {\em et~al.\/}(2020)Murata, Fukami \& Fukagata]{murata2020}
{\sc \au{Murata, T.}, \au{Fukami, K.} \& \au{Fukagata, K.}} \yr{2020}
  \at{Nonlinear mode decomposition with convolutional neural networks for fluid
  dynamics}.  \jt{Journal of Fluid Mechanics}  \bvol{882},  \pg{A13}.

\bibitem[Page {\em et~al.\/}(2021)Page, Brenner \& Kerswell]{Page_2021}
{\sc \au{Page, J.}, \au{Brenner, M.~P.} \& \au{Kerswell, R.~R.}} \yr{2021}
  \at{Revealing the state space of turbulence using machine learning}.
  \jt{Physical Review Fluids}  \bvol{6}~(3).

\bibitem[Platt {\em et~al.\/}(1991)Platt, Sirovich \& Fitzmaurice]{platt1991}
{\sc \au{Platt, N.}, \au{Sirovich, L.} \& \au{Fitzmaurice, N.}} \yr{1991}
  \at{An investigation of chaotic {K}olmogorov flows}.  \jt{Physics of Fluids
  A: Fluid Dynamics}  \bvol{3}.

\bibitem[{Plaut}(2018)]{plaut2018}
{\sc \au{{Plaut}, E.}} \yr{2018}  \at{{From Principal Subspaces to Principal
  Components with Linear Autoencoders}}.  \jt{arXiv e-prints}  \pg{p.
  arXiv:1804.10253},  \arxiv{arXiv: 1804.10253}.

\bibitem[Rowley {\em et~al.\/}(2003)Rowley, Kevrekidis, Marsden \&
  Lust]{Rowley_2003}
{\sc \au{Rowley, C.~W.}, \au{Kevrekidis, I.~G.}, \au{Marsden, J.~E.} \&
  \au{Lust, L.}} \yr{2003}  \at{Reduction and reconstruction for self-similar
  dynamical systems}.  \jt{Nonlinearity}  \bvol{16}~(4),  \pg{1257--1275}.

\bibitem[Sirovich(1987)]{SIROVICH1987}
{\sc \au{Sirovich, L.}} \yr{1987}  \at{Turbulence and the dynamics of coherent
  structures part ii: Symmetries and transformations}.  \jt{Quarterly of
  Applied Mathematics}  \bvol{45}~(3),  \pg{573--582}.

\bibitem[Wan {\em et~al.\/}(2018)Wan, Vlachas, Koumoutsakos \&
  Sapsis]{Wan_2018}
{\sc \au{Wan, Z.~Y.}, \au{Vlachas, P.}, \au{Koumoutsakos, P.} \& \au{Sapsis,
  T.}} \yr{2018}  \at{Data-assisted reduced-order modeling of extreme events in
  complex dynamical systems}.  \jt{PLOS ONE}  \bvol{13}~(5),  \pg{1--22}.

\end{thebibliography}
\end{document}